\documentclass[twocolumn,aps, prb,showpacs,amsmath,amssymb,superscriptaddress, 10pt]{revtex4-1}

\usepackage{times}
\usepackage{graphicx}
\usepackage{dcolumn}
\usepackage{bm}
\usepackage{natbib}

\begin{document}

\title{Determination of the spin-flip time in ferromagnetic SrRuO$_{3}$ from time-resolved Kerr measurements}

\author{C.L.S. Kantner}
\affiliation{Department of Physics, University of California, Berkeley, CA 94720}
\affiliation{Materials Science Division, Lawrence Berkeley National Laboratory, Berkeley, CA 94720}
\author{M.C. Langner}
\affiliation{Department of Physics, University of California, Berkeley, CA 94720}
\affiliation{Materials Science Division, Lawrence Berkeley National Laboratory, Berkeley, CA 94720}
\author{W. Siemons}
\affiliation{Department of Materials Science and Engineering, University of California, Berkeley, CA 94720}
\author{J.L. Blok}
\affiliation{MESA$^{+}$ Institute for Nanotechnology, University of Twente, 7500 AE Enschede, The Netherlands}
\author{G. Koster}
\affiliation{MESA$^{+}$ Institute for Nanotechnology, University of Twente, 7500 AE Enschede, The Netherlands}
\author{A.J.H.M. Rijnders}
\affiliation{MESA$^{+}$ Institute for Nanotechnology, University of Twente, 7500 AE Enschede, The Netherlands}
\author{R. Ramesh}
\affiliation{Department of Physics, University of California, Berkeley, CA 94720}
\affiliation{Department of Materials Science and Engineering, University of California, Berkeley, CA 94720}
\author{J. Orenstein}
\affiliation{Department of Physics, University of California, Berkeley, CA 94720}
\affiliation{Materials Science Division, Lawrence Berkeley National Laboratory, Berkeley, CA 94720}

\date{\today}

\begin{abstract}
We report time-resolved Kerr effect measurements of magnetization dynamics in ferromagnetic SrRuO$_{3}$. We observe that the demagnetization time slows substantially at temperatures within 15K of the Curie temperature, which is $\sim$ 150K. We analyze the data with a phenomenological model that relates the demagnetization time to the spin flip time.  In agreement with our observations the model yields a demagnetization time that is inversely proportional to T-T$_{c}$. We also make a direct comparison of the spin flip rate and the Gilbert damping coefficient showing that their ratio very close to k$_{B}$T$_{c}$, indicating a common origin for these phenomena.
\end{abstract}

\maketitle

\textbf{I: Introduction}

There is increasing interest in controlling magnetism in ferromagnets.  Of particular interest are the related questions of how quickly and by what mechanism the magnetization can be changed by external perturbations.  In addition to advancing our basic understanding of 
magnetism, exploring the speed with which the magnetic state can be changed is crucial to applications such as ultrafast laser-writing 
techniques.  Despite its relevance, the time scale and mechanisms underlying demagnetization are not well understood at a microscopic level.

Before Beaurepaire et al.'s pioneering work on laser-excited Ni in 1996, it was thought that spins would take nanoseconds to rotate, with 
demagnetization resulting from the weak interaction of spins with the lattice.  The experiments on Ni showed that this was not the case and that demagnetization could occur on time scales significantly less than 1 ps\cite{beaurepaire}.  Since then demagnetization is usually attributed to Elliott-Yafet mechanism, in which the rate of electron spin flips is proportional to the momentum scattering rate.  Recently Koopmans et al. have demonstrated that electron-phonon or electron-impurity scattering can be responsible for the wide range of 
demagnetization time scales observed in different materials\cite{koopmansnm}.  Also recently it has been proposed that 
electron-electron scattering should be included as well as a source of Elliott-Yafet spin flipping, and consequently, 
demagnetization\cite{eecorrelation}.  Although Ref. \cite{eecorrelation} specifically refers to interband scattering at high energies, it is plausible that intraband electron scattering can lead to spin memory loss as well.

Time-resolved magneto-optical Kerr effect (TRMOKE) measurements have been demonstrated to be a useful probe of ultrafast laser-induced 
demagnetization\cite{beaurepaire}.  In this paper we report TRMOKE measurements on thin films of SRO/STO(111) between 5 and 165K. Below 
about 80 K we observe damped ferromagnetic resonance (FMR), from which we determine a Gilbert damping parameter consistent with earlier 
measurements on SrTiO$_{3}$ with (001) orientation \cite {langner}. As the the Curie temperature ($\sim 150 K$) is approached the 
demagnetization time slows significantly, as has been observed in other magnetic systems \cite {ogasawara}. The slowing dynamics have been attributed to critical slowing down, due to the similarities between the temperature dependencies of the demagnetization time and the relaxation time\cite{slowingdown}. In this paper we develop an analytical expression relating the demagnetization time to the spin-flip time near the Curie temperature.  This provides a new method of measuring the spin-flip time, which is essential to understanding the dynamics of laser-induced demagnetization.

\textbf{II: Sample Growth and Characterization}

SRO thin films were grown via pulsed laser deposition at 700$^{\circ}$C in 0.3 mbar of oxygen and argon (1:1) on TiO$_{2}$ terminated STO(111)\cite{koster}.  A pressed pellet of SRO was used for the target material and the energy on the target was kept constant at 2.1 J/cm$^{2}$.  High-pressure reflection high-energy electron diffraction (RHEED) was used to monitor the growth speed and crystallinity of the SRO film in situ.  RHEED patterns and atomic force microscopy imaging confirmed the presence of smooth surfaces consisting of atomically flat terraces separated by a single unit cell step (2.2 \AA in the [111] direction).  X-ray diffraction indicated fully epitaxial films and x-ray reflectometry was used to verify film thickness.  Bulk magnetization measurements using a SQUID magnetometer indicated a Curie temperature, T$_{c}$, of $\sim$155K. Electrical transport measurements were performed in the Van der Pauw configuration and show the residual resistance ratio to be about 10 for these films.

\textbf{III: Experimental Methods}
	
In the TRMOKE technique a magnetic sample is excited by the absorption of a pump beam, resulting in a change of polarization angle, $\Delta \Theta_{K}$(t), of a time delayed probe beam.  The ultrashort pulses from a Ti:Sapph laser are used to achieve sub-picosecond time resolution.  Near normal incidence, as in this experiment, $\Delta \Theta_{K}$ is proportional to the $\hat{z}$ component of the perturbed magnetization, $\Delta$M$_{z}$.  $\Delta \Theta_{K}$ is measured via a balanced detection scheme.  For additional sensitivity, the derivative of $\Delta \Theta_{K}${t) with respect to time is measured by locking into the frequency of a small amplitude ($\sim$500 fs) fast scanning delay line in the probe beam path as time is stepped through on another delay line. 

\textbf{IV.1: Experimental Results: Low Temperature}
 
Fig. 1 shows the time derivative of $\Delta \Theta_{K}$ for an 18.5nm SRO/STO(111) sample for the 16ps following excitation by a pump beam, for temperatures between 5 and 85K.  Clear ferromagnetic resonance (FMR) oscillations are present, generated by a sudden shift in easy axis direction upon thermal excitation by a pump beam\cite{langner}.  This motion is described by the Landau-Lifshitz-Gilbert equation with the frequency of oscillation proportional to the strength of the magnetocrystalline anisotropy field, and the damping described by dimensionless phenomenological parameter, $\alpha$.  The motion appears as a decaying oscillation to TRMOKE.  The orientation of the anisotropy field, closer to in-plane with the sample surface in SRO/STO(111) than in SRO/STO(001), makes these oscillations more prominent when observed with the polar Kerr geometry compared to previous measurements.

\begin{figure}
\label{fig:fig1}
\includegraphics[height=2.5in]{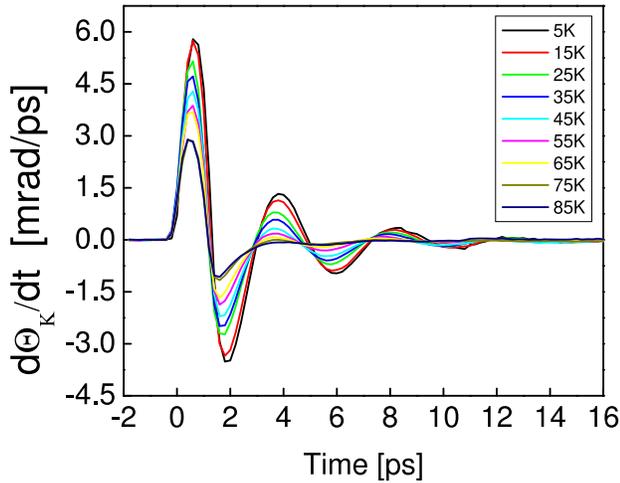}
\caption{Derivative of the change in Kerr rotation as a function of time delay following pulsed photoexcitation, for 5$<$T $<$ 85 K}
\end{figure}

Attempting to model the time derivative of $\Delta \Theta_{K}$ with a damped cosine reveals that it cannot be fit by such a function for t $<$ 2ps.  The feature at short times in Fig. 1 contains higher frequency components, whereas the oscillations which become clear after 2 ps are at a single frequency.   A comparison of the amplitude of the first peak (at t $\sim$.5 ps) with the amplitude of the subsequent oscillations (defined as the difference between d$\Delta \Theta_{K}$/dt at the peak at $\sim$3.5 ps and the dip at $\sim$5.5 ps), is shown as a function of temperature in Fig. 2.  The constant offset between the two amplitudes indicates that d$\Delta \Theta_{K}$/dt is comprised of a superposition of a temperature independent, short-lived component with the longer lived damped oscillations.  

\begin{figure}
\label{fig:fig2}
\includegraphics[height=2.5in]{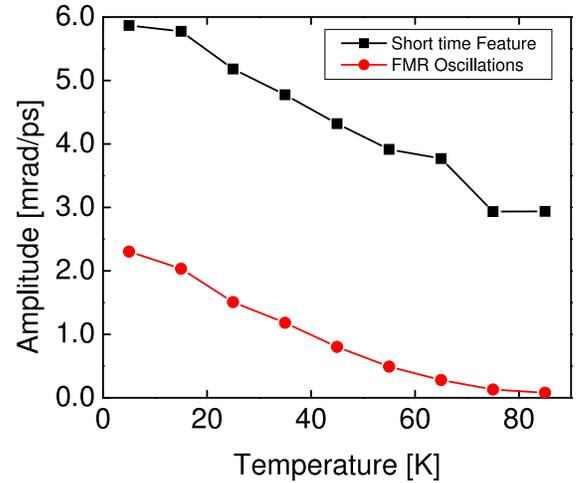}
\caption{Comparing amplitudes of the short time feature and the ferromagnetic resonance oscillations}
\end{figure}

Fitting the oscillatory portion of the signal to a damped cosine, the temperature dependencies of the amplitude, frequency, and damping parameter are found, as shown in Fig. 3.  Comparing these parameters for SRO/STO(111) to previously published work on SRO/STO(001), the frequency is found to be somewhat smaller and to change more with temperature.  Of particular interest is $\alpha$, which is also smaller in this orientation of SRO, consistent with the more pronounced FMR oscillations.  Strikingly, in both orientations there is a dip in $\alpha$ around 45K, which is relatively stronger in SRO/STO(111).  This further strengthens the link between $\alpha$ and the anomalous hall conductivity, speculated in that paper, through near degeneracies in the band structure\cite{langner}.    

\begin{figure}
\includegraphics[height=4.6in]{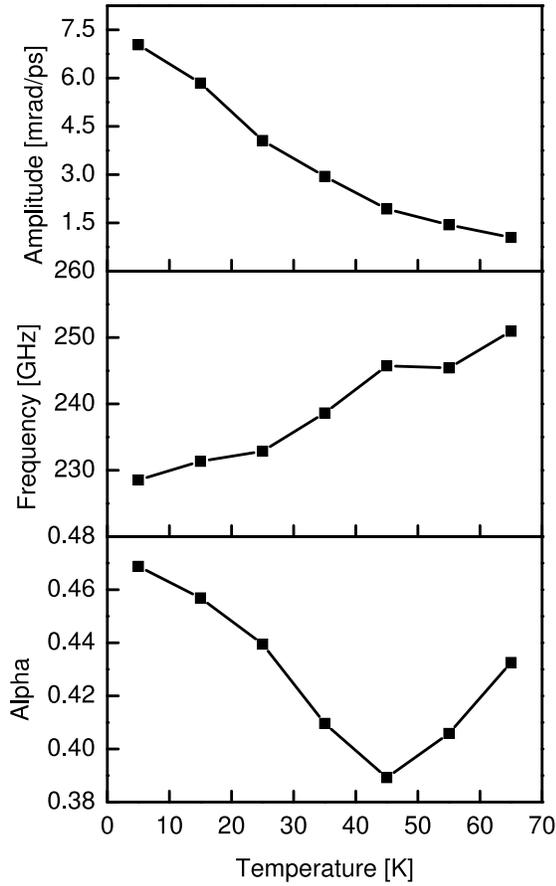}
\caption{Temperature dependence of (a) Amplitude of oscillations, (b) FMR frequency, and, (c) damping parameter}
\end{figure}

\textbf{IV.2: Experimental Results: High Temeperature}

By taking the time derivative of $\Delta \Theta_{K}$, the FMR oscillations can be followed until they disappear at elevated temperatures, at which point it becomes simpler to look at $\Delta \Theta_{K}$ than its time derivative.  Fig. 4 shows $\Delta \Theta_{K}$ as a function of time for the first 38 ps after excitation by the pump laser, for temperatures between 120K and 165K.  A property of a second order phase transitions is that the derivative of the order parameter diverges near the transition temperature.  The peak in magnitude of $\Delta \Theta_{K}$ in figure 4, shown in figure 5, can be understood as the result of the derivative of magnetization with respect to temperature becoming steeper near the Curie temperature.  A strong temperature dependence of the demagnetization time, $\tau_{M}$, is seen, with $\tau_{M}$ significantly enhanced near 150K, consistent with previous reports on SRO\cite{ogasawara, langner}.  

\begin{figure}
\includegraphics[height=3.5in]{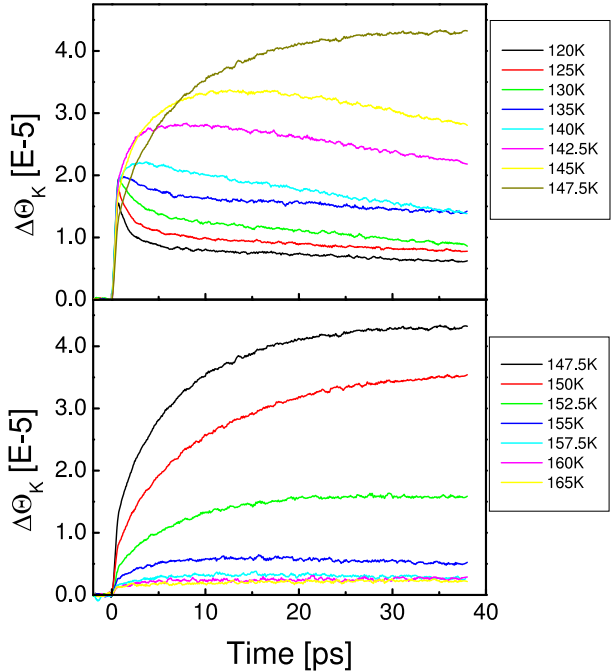}
\caption{Change in Kerr rotation as a function of time delay following pulsed photoexcitation, for 120$<$T $<$ 165 K}
\end{figure}

\begin{figure}[floatfix]
\includegraphics[height=2.5in]{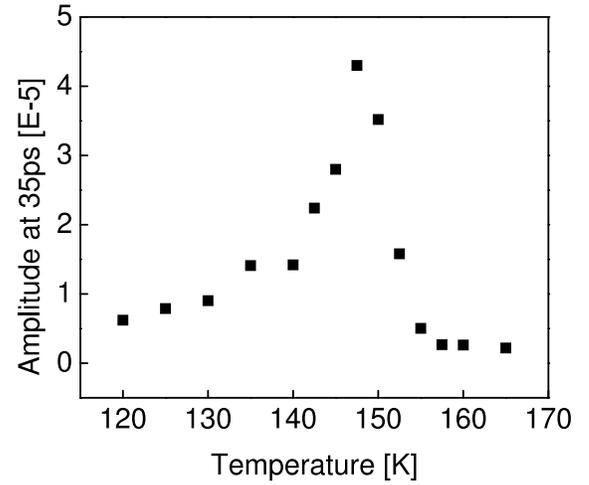}
\caption{Magnitude of change in Kerr rotation at 38ps as a function of temperature}
\end{figure}

$\Delta \Theta_{K}$(t) in Fig. 4, normalized by the largest value of $\Delta \Theta_{K}$(t) in the first 38 ps, can be fit with the following function:

\begin{align}
\text{for} \hspace{0.1in} t<0 \hspace{0.4in} \frac{\Delta \Theta_{K}(t)}{\Delta \Theta_{max}(t)} &= 0  \nonumber \\
\text{for} \hspace{0.1in} t>0 \hspace{0.4in} \frac{\Delta \Theta_{K}(t)}{\Delta \Theta_{max}(t)} &= C - A e^{-t/\tau_{M}}  
\end{align}

where the decay time is $\tau_{M}$.  The resulting $\tau_{M}$ is plotted as a function of temperature in Fig. 6.  Notably, $\tau_{M}$ increases by a factor of 10 from 135K to 150K.  Taking the fit value of T$_{c}$ = 148.8K, as will be discussed later, $\tau_{M}$ is plotted log-log as function of reduced temperature, $t_{R} = (T_{c} - T)/T_{c}$.  The result looks approximately linear, indicating a power law dependence of $\tau_{M}$ on the reduced temperature.

\begin{figure}
\label{fig:fig5}
\includegraphics[height=2.5in]{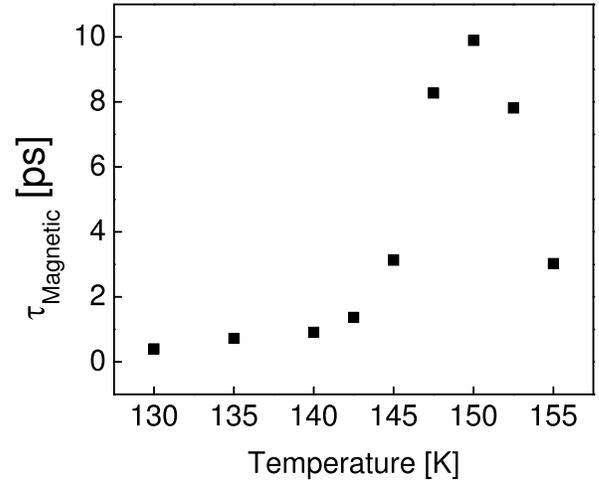}
\caption{Demagnetization time at high temperature}
\end{figure}

\begin{figure}
\label{fig:fig6}
\includegraphics[height=2.5in]{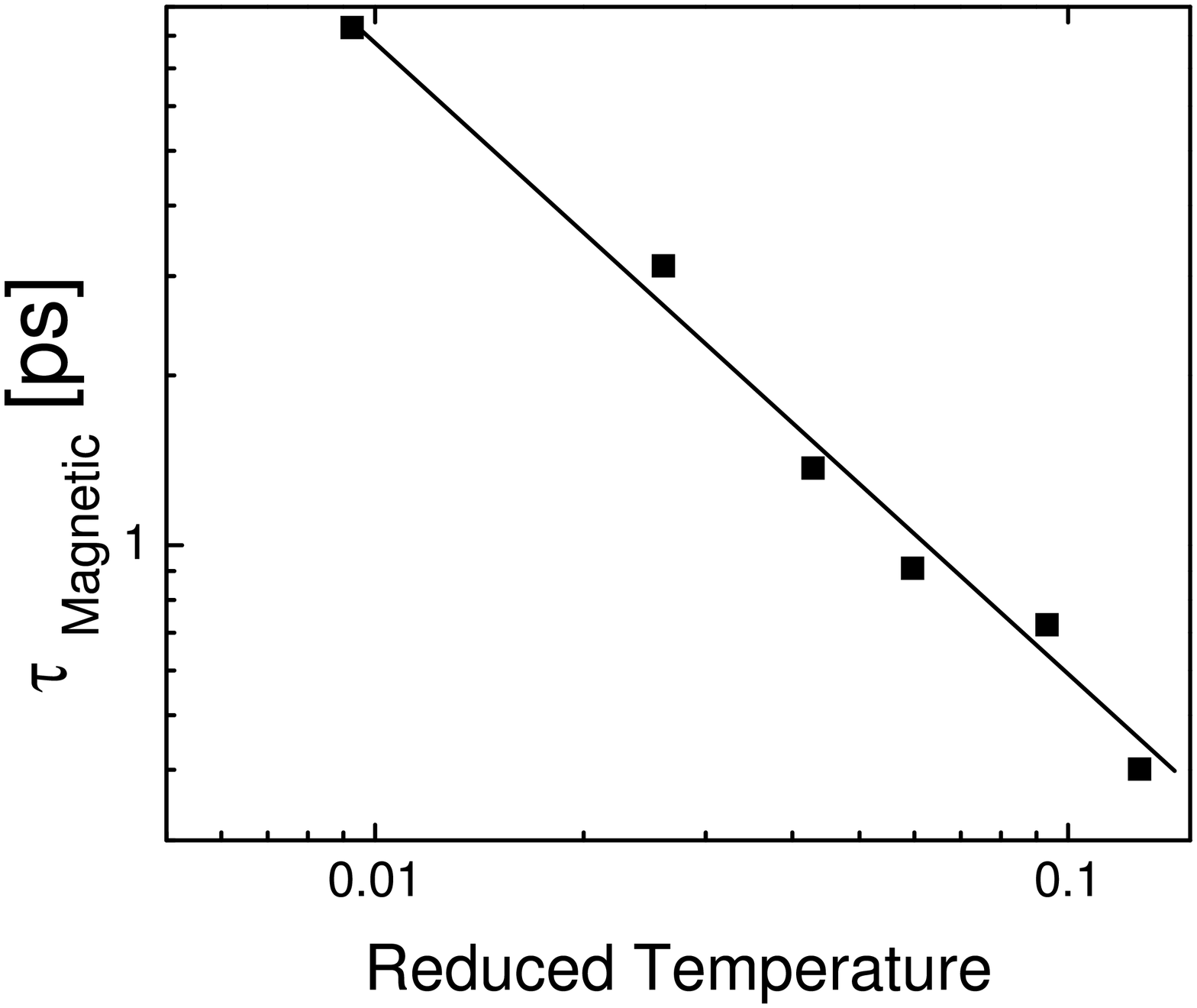}
\caption{Log-log plot of demagnetization time as a function of reduced temperature}
\end{figure}

\textbf{V: Discussion of Results:}

Efforts to explain demagnetization have been largely phenomenological thus far, understandably, given the daunting challenge of a full microscopic model.  Beaurepaire et al. introduced the three temperature model (3TM) to describe demagnetization resulting from the interactions of the electron, phonon, and spin baths\cite{beaurepaire}.  In 3TM the dynamics are determined by the specific heats of each bath as well as the coupling constants between them.  Demagnetization can generally be described with the appropriate choice of coupling constants, providing a guide into the microscopic mechanism.   Koopmans et al. also offer a phenomenological description of demagnetization considering three baths, but one that follows spin in addition to heat\cite{koopmansultrafast}.  Spin is treated as a two state system with energy levels separated by an exchange gap and Fermi's golden rule is used to relate demagnetization to electron scattering which flips a spin.  Equations for coupling constants are derived  based on parameters such as the density of states of electrons, phonons, and spins, the electron-phonon scattering rate, and the probability of spin flip at a scattering event.  

In the following we attempt to understand the behavior of the demagnetization time near T$_{c}$ with an approach based on the two spin state model.  A general relationship between the laser-induced $\tau_{M}$ and the spin flip time, $\tau_{sf}$, can be derived near the transition temperature based on the concept of detailed balance\cite{metropolis}.  In equilibrium, the ratio of  the probability of a spin flipping from majority to minority to the reverse of this process is the Boltzmann factor, $e^{-\Delta_{ex}/kT}$, where $\Delta_{ex}$ is the exchange energy gap.  The time derivative of the number of majority and minority electrons can then be written:

\begin{equation}
\dot{N}_{maj} = - \dot{N}_{min} = \frac{N_{min}}{\tau_{sf}} - \frac{N_{maj}}{\tau_{sf}} e^{-\Delta_{ex}/k_{B} T}
\end{equation}

When the sample is thermally excited by a pump beam, the electron temperature is increased by $\delta T_{e}$.  The rate of change of spins is then altered in the following way:

\begin{equation}
\dot{N}_{maj} = - \dot{N}_{min} = \frac{N_{min}}{\tau_{sf}} - \frac{N_{maj}}{\tau_{sf}} e^{-\Delta_{ex}/k_{B} (T + \delta T_{e})}
\end{equation}

The demagnetization time is related to the total change in spin, $\Delta S$, from initial to final temperature, where, setting $\hbar$=1, $S$ is defined by:

\begin{equation}
S = 1/2(N_{maj} - N_{min})/N_{total}   
\end{equation} 
 
Assuming that $\Delta S$, as a function of time, can be written:

\begin{equation}
\Delta S(t) = [S(T_{f}) - S(T_{i})](1 - e^{-t/\tau_{M}})
\end{equation}
the demagnetization time can be written as:

\begin{equation}
\label{demagtime}
\tau_{M} = \frac{\Delta S}{\dot{S}(0)}
\end{equation}
where $\dot{S}(0)$ is the initial change in the time derivative of the spin.

The total change in spin can be calculated by taking the derivative of $S$ with respect to $T$, and multiplying by $\Delta T_{eq}$, the increase in temperature once electrons, phonons, and spins have come into thermal equilibrium with each other.  $S(T)$ and $\Delta S$ can be written:

\begin{equation}
S(T) = - \frac{1}{2} \hspace{0.05in} tanh \left(\frac{\Delta}{2kT}\right)
\end{equation}
and:

\begin{equation}
\label{koop1}
\Delta S = \frac{d S}{d T} \Big|_{T = T_{0}} \Delta T_{eq} = -\frac{\Delta_{ex}}{4 k_{B} T^{2}_{0}} \left[T_{0} \frac{\Delta'_{ex}}{\Delta_{ex}} - 1\right] \Delta T_{eq}
\end{equation}
where we have relied on the fact that near the transition temperature, $\Delta_{ex} \ll k_{B} T$ and made the approximation that $\delta T_{e} \ll T$ for low laser power.  In the last equation $T_{0} \frac{\Delta'_{ex}}{\Delta_{ex}} \gg 1$ near $T_{c}$, so only the first term will be considered.

The quantity $\dot{S}(0)$, where $\dot{S} = 1/2(\dot{N}_{maj} - \dot{N}_{min})/N_{total}$, can be found by taking the derivative of $\dot{S}(0)$ with respect to $T_{e}$, since immediately after excitation the electron temperature has increased, but the spin temperature, $T$, has not.

\begin{equation}
\label{koop2}
\dot{S}(0) = \frac{d \dot{S}}{d T_{e}} \Big|_{T = T_{0}} \Delta T_{eq} = \frac{N_{maj}}{N_{0} \tau_{sf}} \frac{\Delta_{ex}}{k_{B} T} \Delta T_{eq}
\end{equation}

Near the Curie temperature $N_{maj} \sim N_{min} \sim \frac{1}{2}  N_{total}$.  Using this approximations and equation (\ref{demagtime}), we find: 

\begin{equation}
\tau_{M} = \left(\frac{\Delta'_{ex}}{\Delta_{ex}}\right) \frac{T_{c} \tau_{sf}}{2}  
\end{equation}
where $\Delta_{ex}'$ is the derivative of $\Delta_{ex}$ with respect to temperature and $\Delta_{ex} \sim (T_{c} - T)^{\beta}$, where $\beta$ is the critical exponent of the order parameter.  Taking the derivative, we find 
$\Delta_{ex}' \sim - \beta (T_{c} - T)^{\beta-1}$, and thus can write  

\begin{equation}
\tau_{M} = \frac{\beta \tau_{sf}}{2} \left(\frac{T_{c}}{T_{c} - T}\right)
\end{equation}

Therefore $\tau_{M}$ is predicted to scale as $1/(T_{c} - T)$ near the transition temperature.  A fit of T$_{c} \sim$ 148.8K is found for the data in Fig 6.  

Note that detailed balance suggests that the demagnetization time scales as $1/t_{R}$ near the transition temperature regardless of the underlying mechanism of the demagnetization.  Additionally, the critical exponent found is independent of $\beta$.  It should also be noted that the current situation, where the sample has been excited by a laser, is distinct from critical behavior as typically considered.  In general, divergent time scales are linked to divergent length scales, but here excitations of various length scales are not being excited.  Instead the length scale is always effectively infinite, having been determined by the laser spot size.  $\tau_{sf}$ is plotted as a function of temperature for the mean field value of $\beta = 1/2$, which has been shown to be suitable for SRO\cite{hellman}, in Fig. 8.  $\tau_{sf}$ is revealed to be approximately 200 fs and nearly constant as a function of temperature.

\begin{figure}[floatfix]
\label{fig:fig7}
\includegraphics[height=2.5in]{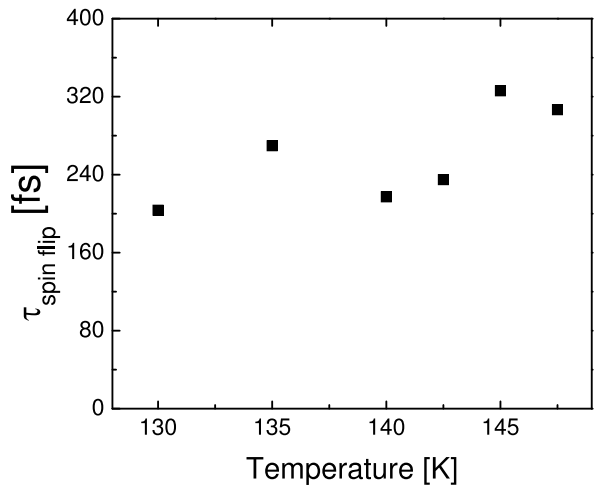}
\caption{Spin flip time at high temperature}
\end{figure}

Previous reports of conductivity in SRO give a scattering time of $\sim$20 fs near the transition temperature\cite{dodge}.  A comparison of the spin flip time with the scattering time implies a probability of 0.1 that a scattering events results in a spin flip.  Though electron-phonon interactions are the most commonly considered source of demagnetization, as mentioned previously, Eliot Yafet-like electron-electron coulomb scattering can also result in demagnetization\cite{eecorrelation}.  This is especially true for materials with strong spin orbit coupling, such as SRO.  Additionally in SRO the interaction with the crystal field means that total spin is not conserved[Goodenough], so every electron interaction can perturb the spin state.   

Having found a relationship between the demagnetization time and the spin flip time we would like to explore the relationship between these parameters and the damping parameter, $\alpha$.  Intuitively, the damping parameter should be proportional to the spin flip scattering rate, or inversely proportional to the spin flip scattering time: $\alpha \sim 1/\tau_{sf}$.  Elliot-Yafet type scattering dissipates energy from motion described by the LLG equation by disrupting the coherent, collective precession of spins.  Spins that have had their angular momentum changed through electron collisions must be pulled back into the precession through the exchange interaction, representing a transfer of energy away from the precessional motion.  These collision-mediated spin-orbit coupling effects are thought to be the primary source of Gilbert-type damping in ferromagnets\cite{heinrich}.  Again, this should be particularly true in a ferromagnet with strong spin orbit coupling.

Combining the spin flip time and the damping parameter with Planck's constant reveals an energy scale, $\mathcal{E}$, given by the condition that:

\begin{equation}
\label{alphatsf}
\frac{1}{\alpha} \sim \frac{\mathcal{E}}{\hbar} \tau_{sf}
\end{equation}

Noting that the values for $\alpha$ and $\tau_{sf}$ found in figures 3 and 7, respectively, are approximately constant as a function of temperature, this energy scale for SRO is $\sim$7 meV.   The fundamental energy scales applicable to the magnetic system in SRO are the Fermi energy, the exchange energy, and the critical temperature, the last two of which are interdependent.  The Fermi energy is orders of magnitude larger than 7 meV, but the energy associated with the critical temperature, $k_{B} T_{c}$ $\sim$ 13 meV, is of the same order.  This suggests an underlying connection between the critical temperature (and thus the exchange energy), Gilbert damping, and spin flip scattering. 

A relationship similar to equation (\ref{alphatsf}) has been found previously between $\tau_{M}$ (rather than $\tau_{sf}$) and $\alpha$ by Koopmans et al. at low temperature:  

\begin{equation}
\tau_{M} = \frac{1}{4} \frac{\hbar}{k_{B} T_{c}} \frac{1}{\alpha}
\end{equation}
   
Applying this equation to SRO at 5K yields $\tau_{m} \sim$ 30fs, which is unphysical since it is below the total scattering rate of $\sim$100fs at low temperature\cite{dodge}.  Whether the fundamental relationship is between transition temperature and the demagnetization time or the spin-flip scattering time remains a question for a microscopic model to resolve. 

\begin{acknowledgments}
This research is supported by the US Department of Energy, Office of Science under contract number DE-AC02-05CH1123.
\end{acknowledgments}

\end{document}